\documentclass[12pt]{iopart}

\usepackage{graphicx}
\usepackage{iopams}
\usepackage{bm}
\usepackage{psfrag}

\newcommand{\be}{\begin{equation}}
\newcommand{\ee}{\end{equation}}
\newcommand{\bea}{\begin{eqnarray}}
\newcommand{\eea}{\end{eqnarray}}

\begin{document}

\title{Interacting electrons on a quantum ring: exact and 
variational approach}

\author{S S Gylfadottir, A Harju, T Jouttenus and C Webb}
\address{
Laboratory of Physics, Helsinki University of Technology, P. O. Box
4100 FIN-02015 HUT, Finland
}
\ead{ssg@fyslab.hut.fi}

\begin{abstract}

We study a system of interacting electrons on a one-dimensional
quantum ring using exact diagonalization and the variational quantum
Monte Carlo method. We examine the accuracy of the Slater-Jastrow
-type many-body wave function and compare energies and pair
distribution functions obtained from the two approaches. Our results show that
this wave function captures most correlation effects. We then study the 
smooth transition to a regime where the electrons localize in the 
rotating frame, which for the ultrathin quantum 
ring system happens at quite high electron density.

\end{abstract}


\submitto{\NJP}

\maketitle


\section{Introduction}

Various low-dimensional electron systems have been the subject of
considerable scientific interest for more than two decades. Examples
of these structures can be formed in a two-dimensional electron gas by
applying appropriate confinement, thus restricting the motion of
electrons to a small area on the boundary between two semiconducting
materials. These systems are in many ways similar to atoms, however,
their properties can be controlled by adjusting their geometry, the
external confinement, and applied magnetic field. Nanostructures are a
source of discoveries of novel quantum phenomena which do not appear
in atoms. They are important both in connection with potential device
applications and can function as convenient samples to probe the
properties of many-electron systems in reduced dimensions.

In quantum rings, electrons are confined to move on a circular region
in space. They have many interesting and actively explored properties: 
Under the influence of a magnetic field, an equilibrium current flows 
along the ring. Furthermore, the energy
spectrum is periodic in the magnetic flux with a period of the flux
quantum $\Phi_0 = h/e$ in case of an even number of noninteracting
electrons, but $\Phi_0/2$ for an odd number. Recently, the existence
of a fractional periodicity has been discovered
both in experiments \cite{Keyser03} as well as in theoretical
and computational studies 
\cite{Kusmartsev91,Jagla93,Kusmartsev94,Niemela96,Deo03,Emperador03,Hallberg04}. 
This is often referred to as the fractional Aharonov-Bohm effect.

In this paper we study the accuracy of a
variational quantum Monte Carlo (VMC) approach, using the
Jastrow-Slater wave function, to quantum ring systems. This approach
has been found to be extremely accurate in the case of quantum
dots \cite{Harju05}. The main advantages of quantum Monte Carlo (QMC)
methods are their applicability to large electron systems and their
accuracy in capturing electron-electron correlation effects. To date,
very few QMC studies of the electronic properties of nanoscale
semiconductor quantum rings have been done. Emperador \etal 
investigated the fractional Aharonov-Bohm effect in few-electron
quantum rings using multi-configurational diffusion Monte Carlo
 \cite{Emperador03} and Borrmann and Harting used path integral Monte
Carlo to study the transition between spin ordered and disordered
Wigner crystal rings \cite{Borrmann01}.
We are not aware of VMC calculations for the quantum rings, and due to
this, even the accuracy of the Jastrow-Slater wave function is an
interesting open question. On the other hand, the good performance of
VMC on two-dimensional quantum dots makes it an appealing technique to
study also systems that are closer to being one
dimensional \cite{rasanen:235307,harju:153101}.
As an ultimate test for VMC in this direction, we consider here the
limit of a purely one-dimensional system.

We calculate using VMC the total energy and pair distribution
function of the quantum ring, in zero magnetic field, containing up to
six interacting electrons. The interaction is taken to be Coulombic.
To test the accuracy of our approach we perform exact diagonalization
(ED) calculations and compare the results. 
We will show that VMC gives a very accurate estimate of the ground
state energy for all the cases studied here. In addition, the charge
part of electron correlations is recovered in very satisfactory
fashion. The only weak point of VMC is that the antiferromagnetic 
nature of the system seen in the exact treatment is not present in 
the VMC results. This shows that even though the simple
VMC wave function is adequate for the charge part, the spin structure
of the Jastrow-Slater wave function is not correct. On the other hand, 
the spin part has only a minor role in, e.g., the total energy.

Already at a moderate strength of interaction, the few-electron quantum ring 
shows a separation of spin and charge degrees of freedom. The spin Hamiltonian is well 
described by the antiferromagnetic Heisenberg model with exchange 
coupling $J$ \cite{koskinen:205323}, which is very small compared to 
the energy of orbital motion. We compare our results to the 
relations for $J$ derived by Fogler and Pivovarov \cite{fogler:195344} 
and show that they are in good agreement. We discuss the localization 
of electrons in relative coordinates and find a smooth 
transition starting already at $1<r_s<2$.

This paper is divided as follows. In section II we discuss our model
of the quantum ring system and go in considerable detail through the
two methods used to calculate its ground state properties, ED and
VMC. Then the results are presented in Section III. Our
conclusions follow in Section IV.

\section{Model and methods}

\subsection{Model}
We model the semiconductor quantum rings in the effective mass
approximation by a many-body Hamiltonian
\begin{equation}
\mathcal{H} = -\frac{\hbar^2}{2 m^*}\sum_{i=1}^N
\frac{\partial^2}{\partial\theta_i^2} + \sum_{i<j}^N V(\theta_{ij})\ ,
\end{equation}
where $\theta_i$ is the angular coordinate of the $i$th electron and 
$V(\theta_{ij})$ is the interaction potential between each pair of 
electrons.  We assume that the ring is extremely narrow so that the
electrons can only move at a certain radius, leading to a purely
one-dimensional ring. The Coulomb interaction
for a ring system is written as
\begin{equation}
V(\theta_{ij}) =
\frac{V_C}{R\sqrt{4\sin^2\left(\theta_{ij}/2\right)
+\mu^2}}\ ,
\end{equation}
where $R$ is the radius of the ring, $V_C$ controls the strength of the 
interaction, and $\mu$ is a small parameter that eliminates the 
singularity at $\theta_{ij}=0$. As the interaction between particles 
at the same position on the ring is $V_C/\mu R$, one can think the 
regularization of the potential to give the ring a width of $\mu R$.

The single-particle eigenstates of the noninteracting case are plane waves
\begin{equation}
\psi_l(\theta)=\exp(-\mathrm{i}l \theta)/\sqrt{2 \pi R}\ ,
\end{equation}
with energies
\begin{equation}
\varepsilon_l=\frac{\hbar^2}{2 m^* R^2} l^2 \ ,
\end{equation}
where $l=0,\pm 1, \pm 2, \dots$ is the angular momentum quantum number. We 
scale the units so that the energy is measured in units of 
$E_0 = h^2/(2m^*\pi^2R^2)$ and length in units of $R$. The only parameter which 
is tuned is the interaction strength, and for the Coulomb interaction this 
can be translated into a change of the ring radius by the formula 
\begin{equation}
V_C = \frac{R}{2a_B^*}\ ,
\end{equation}
where $a_B^*$ is the effective Bohr radius.  The results presented in
this paper are for interaction strength in the range of 1 to 10, which
corresponds to a radius of 21 nm to 209 nm assuming GaAs material
parameters. For comparison, the quantum ring system studied by Keyser
\etal \cite{Keyser03,Keyser02} had a radius of around 150 nm,
which corresponds to an interaction strength of $V_C=7.2$ in our
units. The width of their ring can be estimated to be around 20 nm,
which is in reasonable agreement with the value of $\mu=0.1$ used in
this work. 

In the following, we will occasionally use the one-dimensional 
$r_s$-parameter, which is the average distance between electrons.
In our model, it is directly proportional to the radius of the ring
and thus also to $V_C$
\begin{equation}
\frac{r_s}{a_B^*}=\frac{2 \pi V_C}{N} \ .
\end{equation}

\subsection{Exact diagonalization}

The exact diagonalization method (ED), originally called configuration
interaction, is a systematic scheme to expand the many-particle wave
function using the noninteracting states as a basis.  This method
traces back to the early days of quantum mechanics, to the work of
Hylleraas \cite{Hylleraas28}. The calculation of matrix elements for
the many-body case was originally derived by Slater and
Condon \cite{Slater29,Condon30,Slater31}, and developed further by
L\"owdin \cite{Lowdin55}. The use of the term ED for this can be
seen as an attempt to replace the term configuration interaction with
a normal mathematical term \cite{Oitmaa05}. This term might in some
cases be misleading, as truly exact results are obtained only in the
limit of an infinite basis.

As a simple example to start from, consider a single-particle
Hamiltonian split into two parts $\mathcal{H} = \mathcal{H}_0 +
\mathcal{H}_I$, where the Schr\"odinger equation of the first part is
solvable:
\begin{equation}
\mathcal{H}_0 \phi_i(\mathbf{r}) = \varepsilon_i \phi_i(\mathbf{r}) \ ,
\label{ED_basis}
\end{equation}
and the wave functions $\phi_i(\mathbf{r})$ form a orthonormal
basis. The solution of the full Schr\"odinger equation can be
expanded in this basis as $\psi(\mathbf{r}) = \sum_i \alpha_i
\phi_i(\mathbf{r})$.  Inserting this into the Schr\"odinger equation
\begin{equation}
\mathcal{H} \psi(\mathbf{r}) = E \psi(\mathbf{r}) \ ,
\end{equation}
results in
\begin{equation}
(\mathcal{H}_0 + \mathcal{H}_I) \sum_i \alpha_i \phi_i(\mathbf{r}) = E
\sum_i \alpha_i \phi_i(\mathbf{r}) \ .
\end{equation}
Using \eref{ED_basis}, multiplying with $\phi_j^*$ and integrating gives
\begin{equation}
\sum_i \alpha_i \varepsilon_i \delta_{ij} + \sum_i \alpha_i \int
\phi_j^*(\mathbf{r})\mathcal{H}_I \phi_i(\mathbf{r}) \rmd\mathbf{r}= E
\sum_i \alpha_i \delta_{ij}
\end{equation}
for every $j$. This can be written as a matrix equation
\begin{equation}
(H_0+H_I) \mbox{\boldmath $\alpha$} = E \mbox{\boldmath $\alpha$} \ ,
\end{equation}
where $H_0$ is a diagonal matrix whose $i$th element is
$\varepsilon_i$, and the $ji$th element of $H_I$ is $\int
\phi_j^*(\mathbf{r})\mathcal{H}_I \phi_i(\mathbf{r}) \rmd\mathbf{r}$.
The vector $\mbox{\boldmath $\alpha$}$ contains the values $\alpha_i$.
In this way, the Schr\"odinger equation can be mapped to a matrix form. In
principle, the basis $\{\phi_i\}$ is infinite, but the calculations are
done in a finite basis. The main computational task is to calculate
the matrix elements of $H_I$ and to diagonalize the matrices. The
convergence of the expansion depends on the actual values of the
matrix elements, and in the case where $\mathcal{H}_I$ is only a small
perturbation to $\mathcal{H}_0$ it is fastest.

In a similar fashion, one can split the many-particle Hamiltonian into
two parts, typically a noninteracting and an interacting one. The
solution to the noninteracting many-Fermion problem is a Slater
determinant formed from the $N$ lowest energy eigenstates of the
single-particle Hamiltonian.  The basis typically chosen for the
interacting problem is a Slater determinant basis constructed from $M$
single-particle states. For an $N$-particle
problem, $M\choose N$ 
different many-particle configurations can be
constructed by occupying any $N$ of the $M$ available states. Each of
these determinants is labeled by the $N$ indices of the occupied
single-particle states. Due to interactions, other configurations than
the one of the noninteracting ground state have a finite weight in the
expansion of the many-particle wave function.

The Schr\"odinger equation maps to matrix form in a similar way as in 
the preliminary example above. The calculation of the matrix elements 
of the Hamiltonian are simpler using second quantization and the
occupation number representation. We will only state the results here.  

The noninteracting Hamiltonian matrix $H_0$ in the determinant basis
is diagonal with the $i$th element equal to $E_i=\sum_{j=1}^N
\varepsilon_{i_j}$, where $i_j$ labels the occupied single-particle
state.  The elements of the interaction Hamiltonian matrix can be
found by a straight-forward calculation using the anti-commutation
rules of the creation and annihilation operators. The only non-zero
matrix elements are between configurations that differ at most by two
single-particle occupations. If both of the two configurations have
two single-particle states occupied that are unoccupied in the other
one, the interaction matrix element is
\begin{equation}
\pm(V_{pqrs}-V_{pqsr}) \ ,
\end{equation}
where $p$ and $q$ are single-particle states occupied in the left 
configuration and unoccupied in the right one, and similarly for $s$ 
and $r$ with right and left interchanged. The sign of the matrix element 
depends on the total number of occupations between $p$ and $r$ and 
between $q$ and $s$ due to anti-commutation of the creation and 
annihilation operators. If this number is odd, the sign is minus, 
otherwise plus. For the Coulomb interaction
\begin{equation}
V_{ijkl}= \delta_{s_i,s_k} \delta_{s_j,s_l}
  \int \phi_i^*(\mathbf{r}_1) \phi_j^*(\mathbf{r}_2)
  \frac{1}{r_{12}} \phi_k(\mathbf{r}_1) \phi_l(\mathbf{r}_2) \rmd
   \mathbf{r}_1 \rmd \mathbf{r}_2 \ ,
\label{ED_V}
\end{equation}
and the delta functions result from the spin part summation. In the 
case of a one-dimensional quantum ring this matrix element has a 
closed form
\begin{equation}
V_{ijkl}=\pi^{-1}Q_{i-k-\frac{1}{2}}\left(\mu^2/2+1\right) \ ,
\end{equation}
where $Q(x)$ is the Legendre function which can be written in 
terms of Gauss's hypergeometric function.

If the two configurations have a difference of only one occupation, the 
final result for the interaction matrix elements is 
\begin{equation}
\label{eq:matel}
\pm \sum_{i} (V_{qiri}-V_{qiir}) \ ,
\end{equation}
where $q$ is occupied in the left configuration and unoccupied in the
right one, and for $r$ the other way around. The sum over $i$ is over
orbitals that are occupied in both configurations.  Again the sign of
the matrix element depends on number of occupations between the
differing orbitals. It is easy to see that the matrix elements in 
\eref{eq:matel} vanish for a rotationally symmetric system, as 
a change in only one occupation cannot conserve the total angular 
momentum. 

The diagonal elements of the interaction Hamiltonian matrix are simply 
\begin{equation}
\sum_{i<j}^N (V_{ijij}-V_{ijji}) \ .
\end{equation}

To summarize, the noninteracting part results in a Hamiltonian matrix
which is diagonal, and the interaction Hamiltonian couples
configurations that differ by at most two occupations. Now that we
know the Hamiltonian matrix elements, we can sketch the basic ED
procedure as follows: First solve for $M$ eigenstates of the 
single-particle Hamiltonian $\mathcal{H}_0$. After those have been 
found, calculate the two-particle matrix elements $V_{ijkl}$ of 
\eref{ED_V}. Construct the $M\choose N$ 
$N$-particle 
configurations from $M$ single-particle states. Here configurations 
with wrong symmetry can be rejected. For example, one can pick only 
configurations with certain $z$-component of the total spin or some 
other good quantum number, like the angular momentum in our case. To 
construct the Hamiltonian matrix, 
calculate the interaction matrix elements of the Hamiltonian 
$\mathcal{H}_I$ between the configurations. Construct the diagonal
noninteracting Hamiltonian matrix elements from the single-particle
eigenvalues. Finally, diagonalize the Hamiltonian matrix. The main 
problem of the ED method is the exponential computational
scaling of the method as a function of the number of particles. An 
other problem is the convergence rate as a function of basis size.

To compare the ED and VMC methods for the ring, we study the total energy 
and the pair distribution function of the electrons in the ring. The pair 
distribution function of the system is defined by
\begin{equation}
\label{eq:pcorr_def}
n_{\sigma}({\bf r})n_{\sigma'}({\bf r}') 
  g_{\sigma\sigma'}({\bf r},{\bf r}')
  = \left<\psi\right|\Psi_{\sigma}^{\dagger}({\bf r})
     \Psi_{\sigma'}^{\dagger}({\bf r}')
     \Psi_{\sigma'}({\bf r}')
     \Psi_{\sigma}({\bf r})\left|\psi\right> \ ,
\end{equation}
where $n_{\sigma}({\bf r})$ is the spin $\sigma$ particle density and
$\Psi_{\sigma}^{\dagger}({\bf r})$ and $\Psi_{\sigma}({\bf r})$ are
the field operators. It describes the probability that, given an
electron of spin $\sigma$ at ${\bf r}$, an electron of spin $\sigma'$
is found at ${\bf r}'$. It therefore gives an idea of the strength of
correlation between electrons. The pair distribution function can be
derived in a similar way as the Coulomb matrix elements using the
anti-commutation rules of the creation and annihilation operators.

\subsection{Variational Monte Carlo method}

The QMC methods are among the most accurate ones
for tackling a problem of interacting quantum particles \cite{Foulkes01}.
Often the simplest QMC method, namely, the variational QMC (VMC), is
able to reveal the most important correlation effects. In many quantum
systems, further accuracy in, e.g., the energy is needed, and in these
cases methods such as the diffusion QMC allow one to obtain more
accurate estimates for various observables.

The VMC strategy of solving the ground state of an interacting system
of $N$ particles (defined by the Hamiltonian operator $\mathcal{H}$
and the particle statistics) starts by constructing the variational
many-body wave function $\Psi$. Then, all the observables of the system
can be computed from it. For example, the energy $E$, which is higher than the
ground state energy $E_0$, can be obtained from the high-dimensional
integrals
\begin{equation}
E_0\le E = \frac{\int \Psi^*(\mathbf{R}) \mathcal{H}
\Psi(\mathbf{R}) \rmd \mathbf{R}}{\int |\Psi(\mathbf{R})|^2 \rmd \mathbf{R}}
\ ,
\end{equation}
where $\mathbf{R}$ is a vector containing all the coordinates of
the $N$ particles. If the particles have $d$ degrees of freedom, the
integrals here are $N \times d$ dimensional. As one would like to have
$N$ reasonable large, the integrals are typically so high-dimensional
that these are most efficiently calculated using a Monte Carlo
strategy. This is based on the limit
\begin{equation}
 \lim_{N_R \to \infty} \frac{1}{N_R}\sum_{i=1}^{N_R} f(\mathbf{R}_i)
=
\frac{1}{V} \int_V 
f(\mathbf{R}) \rmd \mathbf{R} \ ,
\end{equation}
where the points $\mathbf{R}_i$ are uniformly
distributed throughout the volume $V$. The error in the evaluation of
the integral decays as $\propto \sigma_f/\sqrt{N_R}$, where $\sigma_f$
is the standard deviation of the function $f$. A typical trick to
reduce the error is to make $\sigma_f$ smaller by dividing the function $f$
by a similar function $g$ (that we assume to integrate to one). Then,
one can rewrite the sum as
\begin{equation}
\frac{1}{N_R}\sum_{i=1}^{N_R} f(\mathbf{R}_i) =
\frac{1}{N_R}\sum_{i=1}^{N_R} g(\mathbf{R}_i)
\frac{f(\mathbf{R}_i)}{g(\mathbf{R}_i)} =
\frac{1}{N_R}\sum_{i=1}^{N_R}
\frac{f(\mathbf{R}_i^g)}{g(\mathbf{R}_i^g)} \ ,
\end{equation}
where now the points $\mathbf{R}_i^g$ are distributed as $g$. The
error in this approach is now smaller, as the function $g$ was chosen 
such that the standard deviation of $f/g$ is smaller than that of $f$.

Now in VMC, one typically generates the random set of $N_R$
coordinates $\{\mathbf{R}_i\}_{i=1}^{N_R}$ so that they are
distributed according to $|\Psi|^2$, and one obtains for the energy
\begin{equation}
E = \frac{1}{N_R}
\sum_{i=1}^{N_R} \frac{\mathcal{H} \Psi(\mathbf{R}_i)}{\Psi(\mathbf{R}_i)} 
= \frac{1}{N_R}\sum_{i=1}^{N_R} E_L \ ,
\end{equation}
where we have defined the local energy $E_L$ to be
\begin{equation}
E_L=\frac{\mathcal{H} \Psi(\mathbf{R}_i)}{\Psi(\mathbf{R}_i)} \ .
\end{equation}
One should note that $E_L$ is a function of the coordinates, and it
defines the energy at this point in space. If the wave function $\Psi$
solves the Schr\"odinger equation, $E_L$ equals the true energy of the
system. From the Monte Carlo point of view, the error in determining
the energy is related to the standard deviation of the local energy,
and for this reason the statistical error in the VMC energy decreases
as the wave function becomes more accurate.

The sampling of the random points $\mathbf{R}_i$ can be done by, e.g.,
using the basic Metropolis algorithm. In that, the ratio of the sampling 
function at the $i$th and the previous step:
$|\Psi(\mathbf{R}_i)|^2/|\Psi(\mathbf{R}_{i-1})|^2$, needs to be 
calculated and the trial step is accepted if the ratio is larger than a 
uniformly distributed random number between zero and one.

A typical, and also the most commonly used VMC trial many-body wave
function, is the one of a Slater-Jastrow type:
\begin{equation}
\Psi = D_{\uparrow} D_{\downarrow} \rme^{\Omega} \ ,
\label{wf}
\end{equation}
where the two first factors are Slater determinants for the two spin
types, and
\begin{equation}
\Omega = \sum_{i<j} j(\theta_{ij}) \ ,
\end{equation}
where $j(\theta_{ij})$ is a Jastrow two-body correlation factor. The
determinants contain $N$ single-particle orbitals, where the ones in
different spin determinants can be the same. These determinants solve
in some cases a noninteracting or a mean-field Schr\"odinger
equation, but in general, one can choose these orbitals rather freely.
This form of a wave function has proved to be very accurate in many
cases \cite{Foulkes01}. In quantum dots, the Jastrow-Slater wave
function captures most correlation effects very
accurately \cite{Harju05}. The only possible exceptions are found in
the fractional quantum Hall effect regime, where only some of the
states can be well approximated by the Jastrow
correlation. Interesting examples of successful cases are the Laughlin
states \cite{Laughlin_13}. 

One can generalize the Jastrow-Slater wave function to contain
more than one determinant per spin type. This leads to an increased
computational complexity, and for this reason it is often
avoided. Another generalization can be to replace the single-particle
coordinates in the determinants by a set of collective coordinates,
e.g., $\mathbf{r}_i\to \mathbf{r}_i-F(\{\mathbf{r_j}\}_{j=1}^N)$. This
also leads to complications, this time in the calculation of the ratio
of the wave functions in the Metropolis algorithm.

The determinants in our VMC trial wave function are constructed from
the noninteracting single-particle states, and for the two-body
Jastrow factor we use here a simple form of
\begin{equation}
j(\theta)= \sum_{i=0}^{10} \alpha_i \cos(\theta)^i \ ,
\label{Jsimple}
\end{equation}
where $\alpha$'s are variational parameters, different for electron
pairs of same and opposite spin type.

An important ingredient in VMC is the optimization of the variational
parameters of the wave function \cite{Harju05}. For an efficient
optimization of the variational parameters, we need the derivative of
the energy with respect to these parameters. Our wave function is
complex-valued, so we need to generalize the result for the real
case \cite{Lin00} to read:
\begin{eqnarray}
\frac{\partial E}{\partial \alpha_k} &=& \frac{\partial}{\partial \alpha_k}
\frac{\int \Psi^{*}\mathcal{H}\Psi \rmd \Theta}{\int |\Psi|^2 \rmd \Theta} \nonumber 
\\ &=& \frac{1}{\int |\Psi|^2 \rmd \Theta} \int \left[ \Psi^{*'}\mathcal{H}\Psi
+ \left( \Psi^{*'}\mathcal{H}\Psi \right)^{\!\!*} \right] \rmd\Theta \nonumber \\ & & - 
\frac{1}{(\int |\Psi|^2 \rmd \Theta)^2} \int \Psi^{*}\mathcal{H}\Psi \rmd 
\Theta 
\int \left[ \Psi^{*}\Psi^{'} 
+ \left(\Psi^{*}\Psi^{'} \right)^{\!\!*} \right] \rmd\Theta \nonumber 
\\ &=& \frac{2}{\int |\Psi|^2 \rmd \Theta} \Re \int |\Psi|^2 \left( \frac{\Psi^{'}}
{\Psi} \right)^{\!\!\!*} \frac{\mathcal{H}\Psi}{\Psi} \rmd\Theta \nonumber \\ & & - 
\frac{2}{(\int |\Psi|^2 \rmd \Theta)^2} \Re \int |\Psi|^2\frac{\Psi^{'}}{\Psi} 
\rmd\Theta \int  |\Psi|^2 \frac{\mathcal{H}\Psi}{\Psi} \rmd \Theta \nonumber  \\ &=&
2\Re \left\langle \left( \frac{\Psi^{'}}
{\Psi} \right)^{\!\!\!*} E_L \right\rangle - 2\Re \left\langle \frac{\Psi^{'}}
{\Psi} \right\rangle \Big\langle E_L  \Big\rangle \ ,
\label{Complex_partial}
\end{eqnarray}
where $\Psi^{'} \equiv \partial \Psi/\partial \alpha_k$ and $\rmd\Theta
\equiv \rmd\theta_1,\ldots, \rmd\theta_N$. In the second line of the
equation we have used the fact that the Hamiltonian $\mathcal{H}$ is
Hermitian and real. The angle brackets denote an average over the
probability distribution $|\Psi|^2$. This notation is also used in the 
following. In our wave function, the parameters $\alpha_i$ are inside an exponential, 
real-valued Jastrow factor:
\begin{equation}
\label{Psi_prop}
\Psi=D_{\uparrow} D_{\downarrow} \rme^{\Omega(\Theta,\boldsymbol{\alpha})} \ ,
\qquad \Omega \in \mathbb{R} \ ,
\end{equation}
where $\Theta \equiv (\theta_1, \ldots, \theta_N)$, $\boldsymbol{\alpha}
\equiv (\alpha_1, \ldots, \alpha_M)$. As a result we have:
\begin{equation}
\frac{\Psi^{'}}{\Psi} = \Omega^{'} \ , \qquad \Omega^{'} \equiv \frac{\partial
  \Omega}{\partial \alpha_k}
\label{J_prime} \ .
\end{equation}
Using \eref{J_prime} we can rewrite \eref{Complex_partial} as
\begin{equation}
\frac{\partial E}{\partial \alpha_k} 
  =2\Re \left[\left\langle \Omega^{'} E_L \right\rangle -  \left\langle  \Omega^{'} 
  \right\rangle \Big\langle E_L  \Big\rangle\right] \ .
\label{local}
\end{equation}

\section{Results}

\subsection{Convergence and configurations}

\begin{figure}
\begin{indented}
\item[]\includegraphics[width=.55\textwidth]{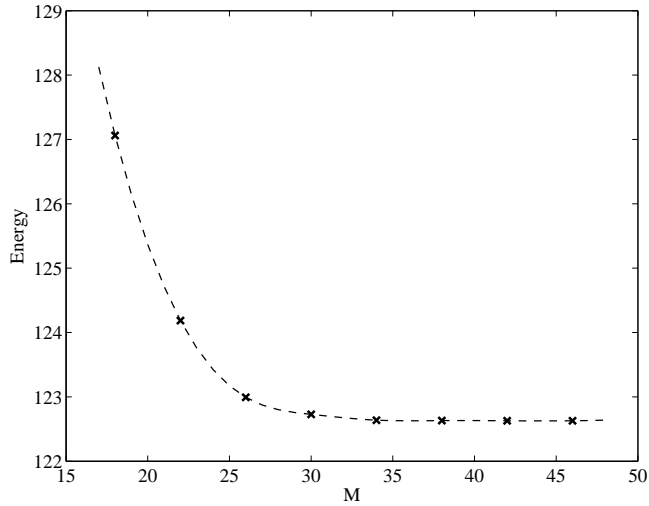}
\end{indented}
\caption{Convergence of the ground state energy as a function of the 
number of single-particle states $M$, for the case of $N=6$ electrons 
and a strong interaction $V_C=10$. The dashed line is to guide the eye.}
\label{fig:conv}
\end{figure}

Before presenting the actual results
we first present an analysis of the accuracy of the ED method
used. The only approximation done in our ED calculations is that we
have a finite number of determinants in our expansion of the
many-electron wave function. The actual number used is finally limited
by the available computer resources. To give an idea of the accuracy
of ED, we show in \fref{fig:conv} the convergence of the $N=6$,
$S=0$ ground state energy with increasing basis size $M$. The
interaction strength is $V_C=10$ making this the worst-case scenario
for the convergence.  We have increased $M$ in steps of four, because
the energy is lowered most when a new pair of $\pm l$ angular momentum
values are included for both spin types. In this case, the maximum
number of basis functions we have used is $M=46$. For these values of
$M$ and $N$, ${M\choose N} = 9\ 366\ 819$. 
However, many configurations
can be rejected from symmetry requirements, as we can force $S_z$ and
$L$ to have certain fixed values. This use of symmetry leads to a
final size of the Hamiltonian matrix of around $79\ 000\times 79\ 000$. 
The difference in energy between $M=42$ and $M=46$ is $4.2\cdot10^{-4}$ 
(in units of $E_0$) 
which is about 0.00034\% of the total energy. We can therefore be 
confident that our ED calculations have sufficient accuracy to be used 
to test the quality of the VMC approach.

\begin{figure}[hbt]
\begin{indented}
\item[]\includegraphics[width=.55\textwidth]{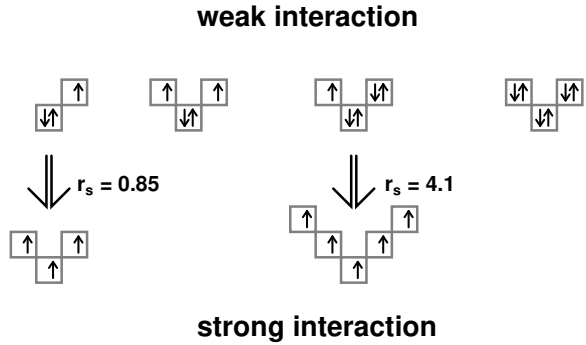}
\end{indented}
\caption{Single-particle states occupied in the VMC wave
function. These are also the most important configurations in the ED
expansion.}
\label{fig:conf}
\end{figure}

When the electrons interact weakly, the many body state has one major 
configuration, that of the equivalent noninteracting system. In this 
limit, the single-determinant Hartree-Fock (HF) method gives reasonable 
results. The electrons arrange so as to minimize the total kinetic energy 
and the single-particle states are occupied as compactly as possible around 
the $l=0$ state. The resulting configurations are shown in the upper 
part of \fref{fig:conf}. 
Each box in this figure represents one single-particle state, having
angular momentum from $l=0$ at the bottom, up through $l=\pm 1, \pm 2,
\dots$, with increasing energy.  For four electrons, the second shell
formed by the angular momentum $l=\pm 1$ states is half-filled, and
the exchange energy favours spin-polarization of the two electrons in
that shell. When the interaction strength grows, the relative importance 
of the kinetic energy decreases with respect to the interaction energy. 
The most important configuration of the many-body wavefunction of strongly 
interacting electrons has closed shells, and thus zero angular 
momentum, for each spin type. The configurations are then either a 
half-filled shell of one spin type or a closed shell. The resulting 
configurations are shown in the lower half of \fref{fig:conf}. For 
$N=4$ and $6$ electrons, the state is already a closed-shell at weak 
interaction and it remains the major configuration in the limit of 
strong interaction. The odd $N$ cases have open-shells at weak interaction 
and make a transition 
%
%
to closed-shell configuration at some point. For a three-electron ring, 
ED calculations show this transition to occur at an interaction strength 
of about $V_C\approx0.41$, corresponding to $r_s\approx0.85\,a_B^*$. 
For five electrons the transition takes place at $V_C\approx3.3$ 
($r_s\approx4.1\,a_B^*$). 
The determinant part of the VMC wavefunction is chosen to be the relevant 
configuration shown in \fref{fig:conf}. These are the main configurations 
of the many-body state found by ED and have, at all except the strongest 
interaction, by far the largest weight in the determinant expansion.
\begin{figure}[hbt]
\begin{indented}
\item[]\includegraphics[width=0.55\textwidth]{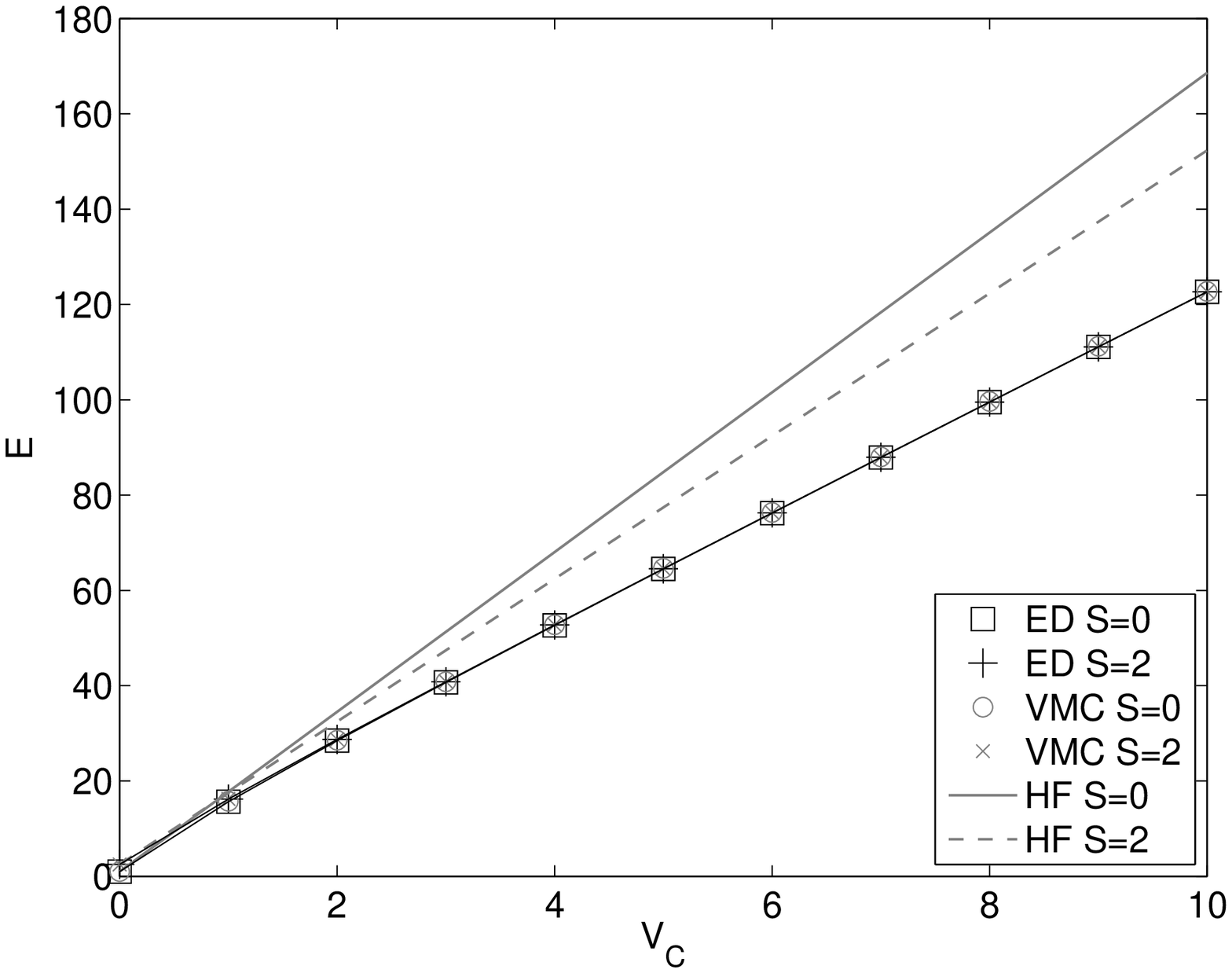}
\end{indented}
\caption{Total energy as a function of interaction strength for 
an $N=6$ electron ring of spin $S=0$ and $S=2$ and angular momentum $L=0$. 
Results from ED, VMC and HF are shown.}
\label{fig:etot}
\end{figure}
\begin{figure}[hbt]
\begin{indented}
\item[]\includegraphics[width=0.55\textwidth]{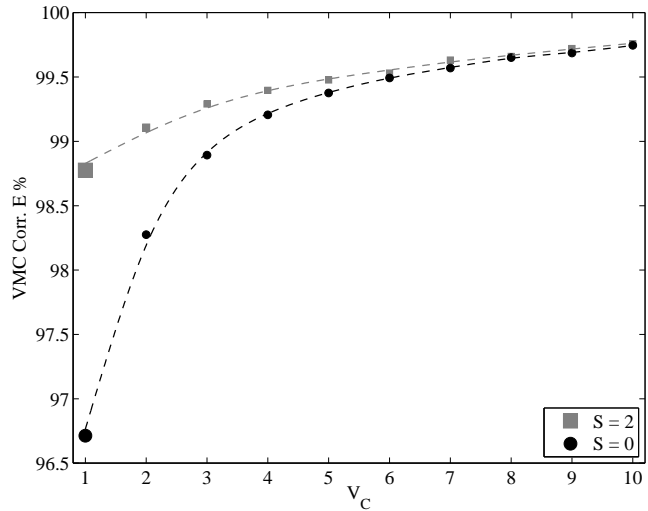}
\end{indented}
\caption{Percentage of correlation energy captured by VMC 
[see \eref{eq:corr_perc} for definition] for $N=6$ electrons, with $L=0$. 
The error in the the first two points is given by the marker size, and is smaller 
than that for the remaining points.}
\label{fig:ecorr}
\end{figure}
\begin{figure*}[hbt]
\begin{indented}
\item[]\includegraphics[width=0.8\textwidth]{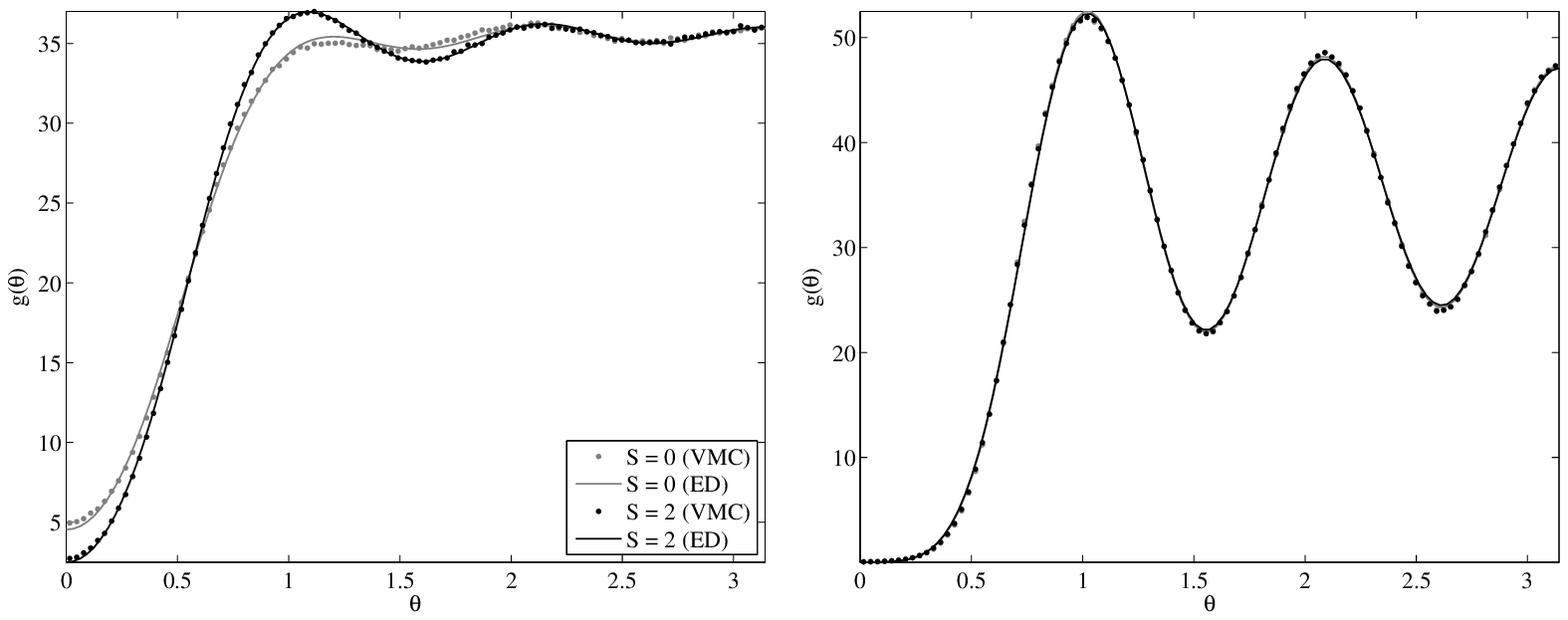}
\end{indented}
\caption{Pair distribution function from ED and VMC for $N = 6$ 
electrons and $L = 0$ with $S=0$ in gray and $S=2$ in black. On the left 
$V_C=1$ and on the right $V_C=10$. The dots show the VMC data while solid lines 
are the ED results. For strong interaction, the curves are nearly identical.
}
\label{fig:pctot}
\end{figure*}

\subsection{VMC accuracy}


The total energy of six electrons on the ring is shown in \fref{fig:etot}. 
We have done the calculations for the $L=0$ state, and a total spin $S=0$ 
(groundstate) and $S=2$, and show the results obtained from ED, VMC and HF 
as a function of the strength of interaction (proportional to the ring radius). 
Over the whole range, ED and VMC results 
agree very well, although the agreement improves slightly with increasing $V_C$. 
The Hartree-Fock results are reasonably accurate only at the lowest interaction 
strength and already at $V_{C}=2$ (corresponding to $R=4\,a_{B}^*$ and 
$r_{s}=2.1\,a_{B}^*$) shows considerable deviation from the ED results.

To compare in more detail the results of ED and VMC,
a convenient measure of the quality of the trial wave function is 
the percentage of correlation energy captured by VMC, defined as
\begin{equation}
\label{eq:corr_perc}
E_C = \frac{E_{\rm HF}-E_{\rm VMC}}{E_{\rm HF}-E_{\rm ED}} \ ,
\end{equation}
where $E_{\rm HF}$ is the Hartree-Fock energy. In \fref{fig:ecorr} we 
show the results for the $L=0$ state of six electrons. At all interaction strengths, 
VMC captures most of the correlation energy. 
It is more accurate for the $S=2$ state than the $S=0$ state, although the 
difference decreases 
with increasing strength of interaction. One would expect VMC to become less accurate 
for strong interactions, because the construction of the VMC wave function 
starts from a single configuration which is later multiplied by a Jastrow factor, 
whereas the many-body wavefunction contains multiple determinants and their 
weights get more evenly distributed with increasing $V_{C}$. This is
not the case, and at the highest $V_{C}$ the deviation in the VMC
groundstate energy lacks such a single-configuration limitation. 
For other particle numbers the results are shown in \tref{tab:e} for low and 
high interaction strength, $V_{C}=1$ and $V_{C}=10$. The VMC results are very 
accurate for the particle numbers we have considered. In the worst case, VMC 
captures 96\% of the correlation energy and the total energy is 
roughly 0.7\% higher than that found by ED. The accuracy improves with 
increasing interaction strength and is better for lower particle numbers. 
This trend is broken by cases where the major configuration has an open 
shell, which is always the case for odd particle numbers at weak interaction. 
This is demonstrated by five electrons, where VMC captures slightly less of 
the correlation energy than for six electrons. 
\begin{table}
\caption{\label{tab:e} Percentages of correlation energy captured by VMC. 
         The error in the last digit is given in parentheses.}
\begin{indented}
\item[]\begin{tabular}{@{}llllll}
\br
$V_C$ & $N=4$     & $N=5$     & $N=6$ \\ 
\mr
1   &   0.9968(2)  & 0.9612(4)  & 0.9671(5)   \\
10  &   0.99909(5) & 0.99917(6) & 0.99746(7) \\
\br
\end{tabular}
\end{indented}
\end{table}
\begin{figure*}[hbt]
\begin{indented}
\item[]\includegraphics[width=0.8\textwidth]{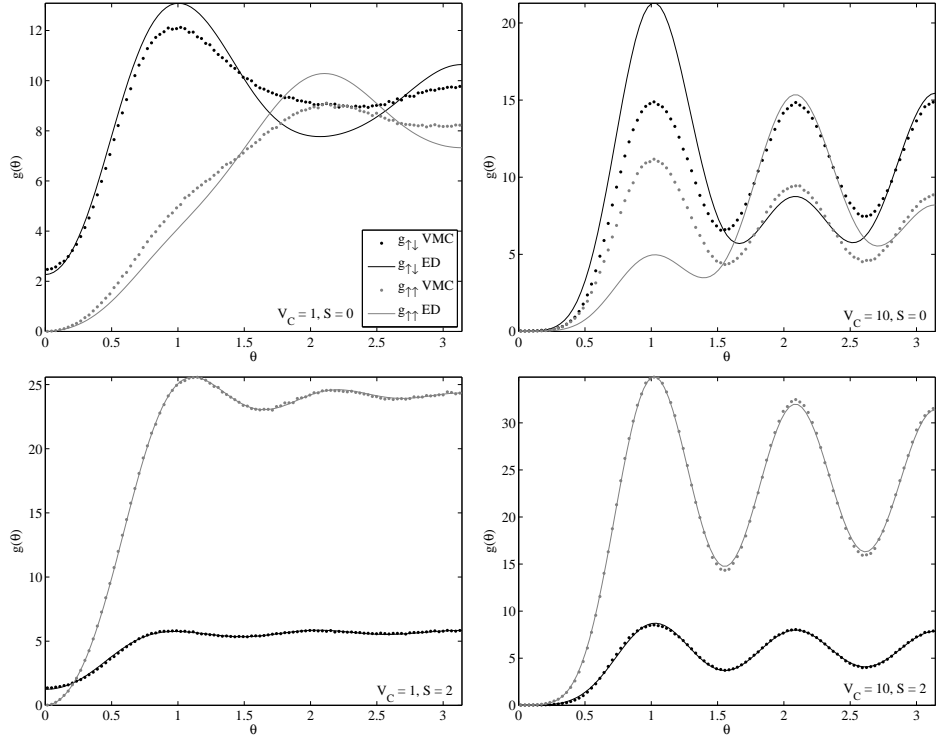}
\end{indented}
\caption{Spin channels in the pair distribution function from ED and VMC for $N = 6$ 
electrons and $L = 0$. The upper panel shows the results for the $S=0$ state and 
interaction strengths $V_C=1$ (left) and $V_C=10$ (right). The lower panel 
shows results for the $S=2$ state $V_C=1$ (left) and $V_C=10$ (right). The 
markers show the VMC data while solid lines are the ED results. 
}
\label{fig:pc_channels}
\end{figure*}
%


In \fref{fig:pctot}, the total pair distribution function of an $N=6$ 
electron ring is shown at weak and strong interaction. As in the case of the 
groundstate energy, the agreement between VMC and ED is good. 
The left panel shows clearly how the total pair distribution functions of different 
spin states become identical when the electrons interact strongly. This is one 
indication of spin-charge separation, when the spin and charge 
of a one-dimensional system decouple. 

\begin{figure}
\begin{indented}
\item[]\includegraphics[width=.55\textwidth]{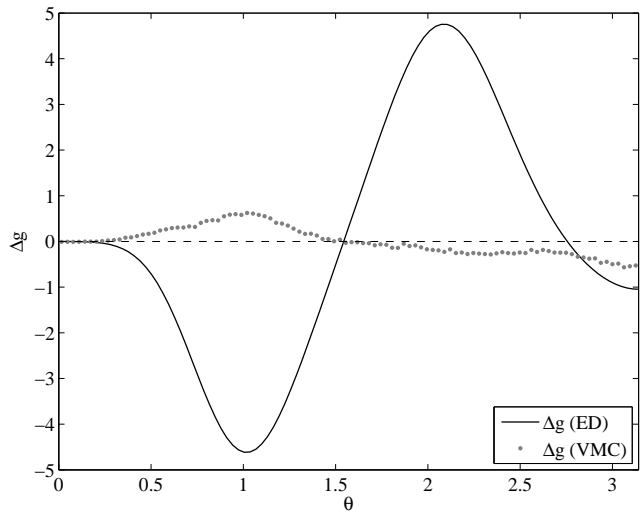}
\end{indented}
\caption{Pair distribution function difference, 
$\Delta g=g_{\uparrow\uparrow}/(N_{\uparrow}-1)-g_{\uparrow\downarrow}/N_{\downarrow}$, 
for six electrons in the state $L=0$, $S=0$. The ED results clearly indicate an 
antiferromagnetic structure.}
\label{fig:gdiff}
\end{figure}

The spin channels in the pair distribution function are shown in
\fref{fig:pc_channels}. For the $S=2$ state, VMC results are
almost identical to those from ED. In the case of $S=0$, however, the
internal structure is inaccurate at weak interaction and completely
wrong at strong interaction. This is further shown in
\fref{fig:gdiff} where we have plotted
\begin{equation}
\Delta g = \frac{g_{\uparrow\uparrow}}{N_{\uparrow}-1}
      -\frac{g_{\uparrow\downarrow}}{N_{\downarrow}}\,,
\end{equation}
which for an antiferromagnetic alignment of spins should give equidistant 
peaks (up spins) and 
troughs (down spins). This is indeed the case, and we furthermore see a 
decay in spin correlations with increasing interelectron distance. The 
VMC results show no antiferromagnetic structure.  
We believe the reason for VMC not being able to handle antiferromagnetic 
spin coupling is due to three-particle correlation which is missing in the 
VMC wavefunction. Despite VMC's shortcomings in accurately 
describing the spin correlations for an antiferromagnetic spin structure, the 
total pair distribution function and the groundstate energy are correctly 
described.


\subsection{Spin and charge in the Wigner crystal regime}

\begin{figure}
\begin{indented}
\item[]\includegraphics[width=.55\textwidth]{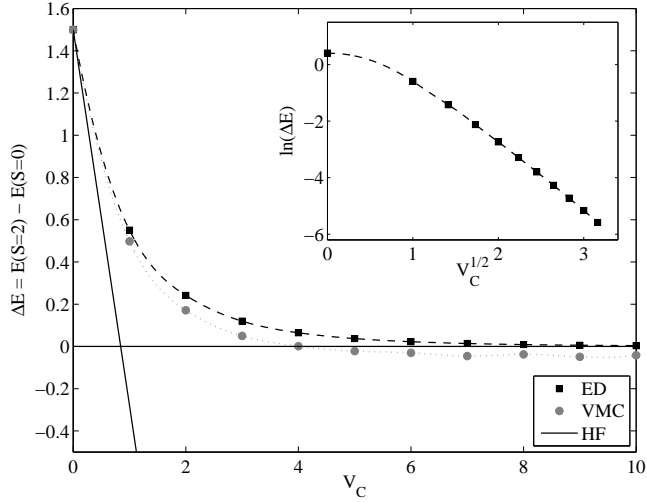}
\end{indented}
\caption{Difference in energy of the $S=0$ and $S=2$ states of six
electrons with angular momentum $L=0$, as found by ED, VMC and HF. In
the inset the logarithm of the energy difference is plotted against
$\sqrt{V_C}$.  }
\label{fig:ediff}
\end{figure}

Like the pair distribution function, the total energy of the $S=0$ 
and $S=2$ states becomes identical in the limit of strong interaction. 
In \fref{fig:ediff} we show the energy difference of these states for six 
electrons, as a function of $V_C$ found by ED, VMC and HF. The inset shows 
the energy difference found by ED on a logarithmic scale. 
As expected, HF fails seriously already 
at low values of $V_C$ and predicts a groundstate of nonzero spin-polarization 
at $V_C$ just below 1 ($r_{s}\approx 1$). VMC follows the ED results more closely, 
but at $V_C=4$ predicts a change in spin. ED results indicate, however, that the 
groundstate spin remains $S=0$ over the whole range of $V_C$ although the energy 
difference approaches zero. Interestingly, the difference between the ED and VMC results 
remains approximately constant for $V_C=4\dots10$.

\Fref{fig:ediff} shows that in the strong interaction regime, the
energy scale of spin dynamics becomes very small when compared to the
energy scale of the orbital motion. This is one indication of a strong
spin-charge separation in the system. It is further manifested in the
pair distribution function, as shown in \fref{fig:pctot}, which is
the same for both spin states, $S=0$ and $S=2$, at stronger
interaction. The spin part can be described by the antiferromagnetic
Heisenberg model \cite{koskinen:205323} with spin interaction 
\be
H_{\sigma}=J\sum_{i,j} {\bf S}_i\cdot{\bf S}_j 
\ee 
with a nearest-neighbour coupling constant $J$. In a recent
paper, Fogler and Pivovarov \cite{fogler:195344} derived a relation
between the coupling constant of thin quantum rings and the pair
distribution function of a spin-polarized 1D system $g_{0}(x)$
\begin{equation}
J=\frac{\rme^2a_{B}^2}{2\ln(a_{B}/R^*)\epsilon}\,g_{0}''(0)\,,
\end{equation}
where $R^*$ is the thickness of the ring, in our case $\mu R$. For
ultrathin wires exhibiting strong spin-charge separation, this
relation holds even at moderate electron densities, $r_{s}\sim 1$.
They furthermore derive an expression for the pair distribution
function, which leads to
\begin{equation}
\label{eq:fogler}
  J = \frac{\kappa}{(2r_{s})^{5/4}}\frac{\pi}{\ln(a_{B}/R^*)}
          \frac{\rme^2}{\epsilon a_{B}}\exp(-\eta\sqrt{2r_{s}}) \ ,
\end{equation}
where $\kappa$ and $\eta$ are constants. 
\begin{figure}[hbt]
\begin{indented}
\item[]\includegraphics[width=.55\textwidth]{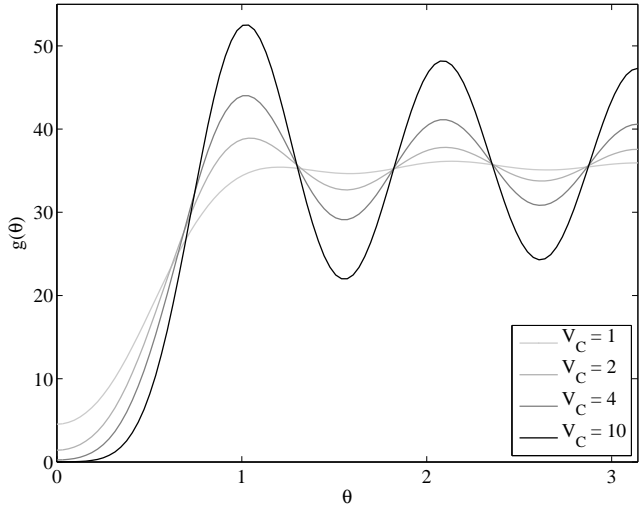}
\end{indented}
\caption{Pair distribution function of six electrons in the $L=0$, $S=0$ state 
for several values of the strength of interaction $V_C$.}
\label{fig:wigner}
\end{figure}
\begin{figure}[hbt]
\begin{indented}
\item[]\includegraphics[width=.55\textwidth]{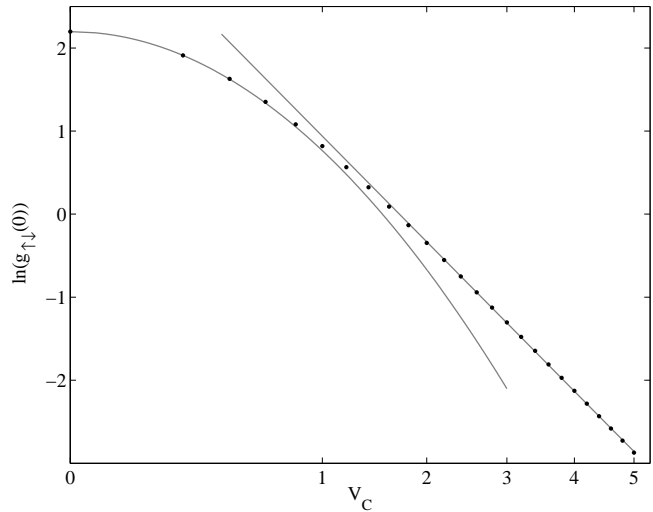}
\end{indented}
\caption{Logarithm of the contact probability $g_{\uparrow\downarrow}(0)$ 
of opposite spin electrons as a function of the strength of interaction 
$V_C$ on a square root scale. The parabolic curve is a guide to the eye 
only and indicates an exponential falloff with $V_C$. The straight line 
is a linear fit and indicates a $\exp(-\zeta\sqrt{V_C})$ behaviour.}
\label{fig:transition}
\end{figure}
The coupling constant $J$ is proportional to the difference in energy
of the $S=0$ and $S=2$ states, $\Delta E$. In our model, $r_{s}=2\pi
V_{C}/N$ so we can write
\begin{equation}
\Delta E\propto\exp(-\eta\sqrt{4\pi/N}\sqrt{V_{C}})\ .
\end{equation}
In the inset of \fref{fig:ediff} we have plotted the logarithm of
$\Delta E$ as a function of $\sqrt{V_C}$. Our results agree very well
with the results of Fogler et al., as the curve becomes linear even at
low values of $V_C$, $V_C\approx 1.7$ ($r_s\approx 1.8$). From our
data we can estimate the slope, and find that $\eta\approx2.5$ which
is reasonably close to the value 2.7979 found by Fogler and
Pivarov. However, it must be noted that our results are for a fixed
value of $\mu=0.1$, so that the thickness of the ring scales linearly
with increasing interaction strength as $R^*/a_{B}^*=0.1R=0.2\,V_{C}$.

The pair distribution function of six electrons of zero total angular
momentum and spin is shown in \fref{fig:wigner} for several values
of $V_C$. 
At $V_C=1$ the distribution is rather flat, but already at
$V_C=2$ there is a clear peak structure indicating partial
localization in relative coordinates. As the system is finite, a 
phase transition to a Wigner crystal does not take place. However, 
there is a smooth transition to a regime where a Wigner-crystal-like 
behaviour is found. The system is translationally invariant and the 
electron density remains uniform, the localization occurs in the 
relative frame.
In two-dimensional 
quantum dots, the electrons are found
to be localized at \cite{egger:3320} $r_s=4$, and the transition is
preceded by a change in the groundstate spin \cite{harju:075309}. A
QMC study of the three-dimensional electron gas showed an abrupt change in
the evolution of the correlation energy with
density \cite{drummond:085116}. We do not see such a change in our
system. One should also note that the three-dimensional QMC work has 
different wave functions for the liquid and solid phase, whereas we
use the same wave function in both cases.

Further insight is obtained from the value of the pair distribution function at
$\theta=0$, which is the contact probability. At low values of the 
strength of interaction $V_C$ it has a nonzero value, as electrons
of opposite spin are not restricted by the Pauli exclusion
principle. As $V_C$ grows, the strong Coulomb repulsion reduces
$g_{\uparrow\downarrow}(0)$ to the point where the electrons become
localized between their neighbors on either side.  A closer look at
the contact probability reveals an indication that the quantum ring
system undergoes a smooth transition at densities between $1<
r_s<2$. In \fref{fig:transition} the logarithm of the contact
probability is shown as a function of the interaction strength on a
square root scale. For $V_C\geq2$, the linear fit indicates that 
$g_{\uparrow\downarrow}(0)$ falls off as $\exp(-\zeta\sqrt{V_C})$, 
where $\zeta$ is a constant. At low interaction strength, the behaviour 
is closer to $\exp(-\zeta'V_C)$, the parabolic curve shown. A related 
behaviour has been found for a different quantity in quantum dots \cite{egger:3320}.

\section{Conclusions}

We have studied interacting electrons in a one-dimensional quantum ring. 
We compared the results 
obtained from variational Monte Carlo with those of exact diagonalization.
The variational wave function was taken to be of a
simple Slater-Jastrow form: a single determinant, the ground state of 
the noninteracting system, multiplied by a two-body correlation factor. 
The results obtained indicate that this simple wave function is able to capture
the important correlation effects, and in most cases even with a high
accuracy.

The major source of error in VMC is the spin energy scale and spin
correlation. The spin energy scale, however, becomes very small
already at rather weak interaction and the error in the groundstate
energy is correspondingly small and decreases with increasing $V_{C}$.
For an antiferromagnetic spin correlation, VMC does not capture the
internal spin structure, but the total pair distribution function is
accurate. From this, one can conclude that the Jastrow factor is very
efficient in capturing the total energy and essential
correlation effects induced by the interaction between electrons. One
might question the importance of the determinant part of the VMC wave
function. We believe that having a good choice for this is crucial for
the success of the VMC simulation. Even if the weight of this single
configuration is not large in the exact expansion when the interaction
is strong, it specifies the symmetry of the VMC many-body state. One
should note that this is also the case for quantum dots, where the
noninteracting single-particle states can be used in the
determinant part of the VMC wave function, and they are even the
optimal ones in the case of a closed-shell
configuration \cite{Harju05}. The reason, both in quantum rings and
dots, is the high symmetry of the problem: In a parabolic dot, the
centre-of-mass and relative motion decouple, also in the many-body
wave function. Thus if one starts from a noninteracting state that is
a ground state of the centre-of-mass motion, the interactions only
change the relative motion part of the many-body wave function. This
means that if a Slater-Jastrow wave function is used, the Slater
determinant part is not changed by the interaction between
electrons. The case of a quantum ring is similar in the sense that the
centre-of-mass and relative motion also decouple, but now because of
the translational invariance of the system.

For a moderate strength of interaction, the spin correlation is well 
described by an antiferromagnetic Heisenberg model with coupling constant 
$J$. We showed that our results agree well with the expression for $J$ derived 
by Fogler and Pivovarov \cite{fogler:195344} at $r_s\gtrapprox 2$. Their result 
is applicable already at $r_s\sim 1$ if a strong spin-charge 
separation is present in the system. This indicates that for an ultrathin quantum ring 
this separation does take place at a quite high electron density. The contact probability 
also shows that there is a change in the properties of the system starting 
at $1<r_s<2$.

In conclusion, a simple variational wave function, combined with
Monte Carlo techniques to calculate the optimal wave function and 
observables of the system, results in an efficient computational strategy for
interacting electrons in purely one-dimensional quantum rings. Since a
similar conclusion can be drawn for two-dimensional quantum
dots \cite{Harju05}, we expect this to hold also for quantum rings of a finite
width.

\ack
This work has been supported by the Academy of Finland through its
Centres of Excellence Program (2000-2005). SSG acknowledges financial 
support from the Vilho, Yrj\"o and Kalle V\"ais\"al\"a Foundation of the 
Finnish Academy of Science and Letters.\\

\bibliographystyle{iopart-num.bst}
\bibliography{Refs}


\end{document}